\documentstyle[psfig,aps,prl]{revtex}
\begin{document}
\title{Quantum Statistical Thermodynamics of Two-Level Systems}
\author{Paul B. Slater}
\address{Community and Organization Research Institute,
University of California, Santa Barbara, CA 93106-2150\\
e-mail: slater@itp.ucsb.edu,
FAX: (805) 893-2790}

\date{\today}

\draft
\maketitle
\vskip -0.1cm

\begin{abstract}
We study four distinct families of Gibbs canonical distributions defined on the
 standard complex, quaternionic, real and  classical/nonquantum
 two-level systems.
The structure function or density of states for any
system is a simple power (1, 3, 0 or -1) of the length of its
 polarization
vector, while the magnitude of the energy of the system, in all four cases,
is the negative of the logarithm of the determinant
of the corresponding
 $2 \times 2$ density matrix. Functional relationships --- proportional to
ratios of gamma functions --- are found between
the average polarizations with respect to the Gibbs
distributions and the
 {\it effective polarization} temperature
parameters. In the standard complex case, this yields an interesting
alternative, meeting certain probabilistic requirements recently
set forth by Lavenda,
to the more conventional (hyperbolic tangent)
 Brillouin function of paramagnetism
(which, Lavenda argues, fails to meet such specifications).
\end{abstract}


%\vspace{-0.4cm}

\section{Introduction}
In a number of forcefully-written papers some twenty years ago, Band and Park
 \cite{bp1,bp2,bp3,bp4} strongly recommended
that the basic objective of quantum statistical thermodynamics should be taken
to be the estimation of a Gibbs distribution (satisfying imposed constraints
on expectation values of observables) over the continuum or ``logical
spectrum'' of possible
density matrices describing the system in question. This was viewed as
a  conceptually preferable alternative to that of
 estimating such a distribution over
simply a set of eigenstates --- corresponding
 to a {\it single} canonical
density matrix --- as in the standard
Jaynesian approach \cite{jaynes,balian}.
They, however, encountered ``two barriers'' in
attempting to establish that their suggested methodology would yield
 identically
the same expectation values as the empirically successful Jaynesian strategy.
``One is essentially philosophical: The prior distribution needed in this
continuum problem because of the inadequacy of the Laplacian rule of
indifference, remains unknown. The other obstacle is mathematical:
$\dots$ we could not perform the required integrals because we have no useful
analytical description of the domain [of density matrices] and its
boundary'' \cite[p. 235]{bp3}. More recently, Park \cite{park}
(cf. \cite{gyft}), in
studying Schr\"{o}dinger's probability relations, wrote that the
``details of quantum thermodynamics are presently unknown'' and
``perhaps there is more to the concept of thermodynamic equilibrium
than can be captured in the canonical density operator itself.''

 The first and principal model considered here
(sec~\ref{sc}) is a particular implementation for the
case of two-level (standard complex) quantum systems of the
general program (formulated in terms of systems of {\it arbitrary}
dimensionality) of Band and Park (cf. \cite{slater}).
It yields 
  seemingly novel analyses of both the thermodynamic properties of a
plane radiation field \cite{bross} and
 of paramagnetic
phenomena (cf.
\cite{tusz,schlogl,gold,bal2}). A noteworthy aspect of our analyses is that
we find it necessary to identify the parameter
$\beta$ of the Gibbs distributions introduced below, not with the inverse
thermodynamic temperature, as is conventional, but rather with
the {\it effective polarization} temperature \cite{bross,christ}.
Asymptotically, we find a {\it negative} log-linear relationship
(equation (\ref{xbetalog}) and Fig.~\ref{plot6})
 between this temperature and the reduced thermodynamic
temperature of the standard model of paramagnetism.

The usual textbook treatment of the particular (spin-1/2) case of
 paramagnetism in which an ensemble
of $n$ noninteracting {\it two}-level systems is subjected
to a magnetic field $H$ predicts that the equilibrium magnetization
($M$) is given by the corresponding (hyperbolic tangent) Brillouin function,
 \cite[p. 192]{lav3},
\begin{equation} \label{brill}
M_{0} \tanh {\mu H \over k T},
\end{equation}
where $M_{0}$ is the saturation magnetization, $\mu$ is the Bohr
magneton, $k$ is Boltzmann's constant and $T$ is temperature.
Brillouin (generalized Langevin)
 functions, in general, it appears, yield good predictions
except for low temperatures and high fields \cite{abrahams}.
In volume I of his text, Balian \cite[sec. 1.4.3]{bal2} writes that
the hyperbolic tangent Brillouin ``model gives a
 qualitative understanding of the saturation
 effect...However, the quantitative agreement is not good as the
behaviour predicted...altogether does not have the same shape as the
experimental curves'' (cf. \cite[Fig. 6.5]{stanley}).

Lavenda \cite[p. 193]{lav3} has recently argued
 that the hyperbolic tangent ``Brillouin function has to coincide with the
first moment of the distribution [for a two-level system having probabilities
${e^{x} \over e^{x} +e^{-x}}$ and ${e^{-x} \over e^{x} + e^{-x}}$],
 and this means that the generating
function is $Z(x)=\cosh x$ [where $x= {\mu H \over k T}$]. Now, it will
be appreciated that this function cannot be written as a definite
 integral, such as
\begin{equation} \label{gener}
Z(\beta) = {1 \over 2} \int_{-1}^{1} e^{\beta x} dx
 = {\sinh \beta \over \beta} =({\pi \over 2 \beta})^{1/2} I_{1/2}(\beta),
\end{equation}
 because the integral form for the hyperbolic
Bessel function,
\begin{equation} \label{bessel}
I_{\nu}(x)= {(x/2)^{\nu} \over \sqrt{\pi} \Gamma({1 \over 2} + \nu)}
\int_{-1}^{1} {e^{\pm x t} \sin^{\nu-1/2} t dt}
\end{equation}
 exists only for $\nu > {1 \over 2}$.
This means that $I_{-1/2}(x) = (2/{\pi x})^{1/2} \cosh x$
cannot be expressed in the above integral form. Since the generating function cannot be derived as the Laplace
transform of a prior probability density, it casts serious doubts on
the probabilistic foundations of the Brillouin function. In other words,
any putative expression for the generating function must be
compatible with the underlying probabilistic structure; that is, it must
be able to be represented as the Laplace transform of a prior
probability density.''

Additionally, as the concluding paragraph of his recent book, Lavenda writes
\cite[p. 198]{lav3}, ``Even in this simple case of the Langevin function,
\begin{equation} \label{langevin}
M= M_{0} \{ \coth({\mu H \over k T}) -{k T \over \mu H} \},
\end{equation}
we have witnessed a transition from a statistics dictated by the
central-limit theorem, at weak-fields, to one governed by extreme-value 
distributions, at strong-fields. Such richness is not possessed by the
Brillouin function, for although it is almost identical to the Langevin
function in the weak-field limit, the Brillouin function becomes
independent of the field in the strong-field limit.
In the latter limit, it would imply complete saturation which does not
lead to any probability distribution. This is yet another inadequacy of
modeling ferromagnetism by a Brillouin function, in the mean field
approximation. And it  is another illustration of our recurring theme that
\textbf{physical phenomena are always connected to probability
distributions}. If the latter does [{\it sic}] not exist, the former is
[{\it sic}] almost
sure to be illusory.''

We derive  here standard complex, quaternionic, real and classical/nonquantum
counterparts --- (\ref{expO}),
 (\ref{quatpolar}), (\ref{realexp}) and (\ref{polKMB}) ---  to the
 hyperbolic tangent
 Brillouin 
function (\ref{brill}), which
 are, evidently, not subject to the objections to it
 that Lavenda has raised.
The question of whether or not they are superior in explaining
physical phenomena, warrants further investigation.

\section{Analyses of Four Types of Two-Level Systems} \label{Analyses}
\subsection{The standard complex case} \label{sc}
\subsubsection{Gibbs canonical distributions}
We consider the two-level quantum systems,
describable (in the standard complex case)
 by $2 \times 2$ density matrices of the form,
\begin{equation} \label{1}
\rho = {1 \over 2}
\pmatrix{\ 1+z&x-iy\cr
        x+iy&1-z\cr},\qquad (x^2+y^2+z^2 \leq 1).
\end{equation}
We assign to such a system
 an energy $(E)$ equal (in appropriate units) to the negative
of the logarithm of the determinant of $\rho$, that is
(transforming to spherical coordinates --- $r, \theta, \phi$ --- so that
$x =r \cos \phi \sin \theta, y = r \sin \phi \sin \theta$ and
$z = r \cos \theta$),
\begin{equation} \label{2}
E = - \log (1-r^2)
\end{equation}
(cf. \cite{tusz}). 
The radial coordinate ($r$) represents the degree of purity
(the length of the polarization vector) of the associated two-level
system. The pure states --- $r=1$ --- are assigned $E=\infty$,
while the fully mixed state --- $r=0$ --- receives $E=0$.
We, then, have (\ref{2}),
\begin{equation} \label{invert}
r = \pm \sqrt{1-e^{-E}}.
\end{equation}
We employ the {\it positive} branch of this solution as
 the density of states or structure function,
\begin{equation} \label{3}
\Omega (E) = \sqrt {1 - e^{-E}} \qquad (0 \leq E \leq \infty).
\end{equation}
(For small values of $E$, this is approximately $\sqrt{E}$.)
The integrated density of states \cite{brezin} is, then,
\begin{equation} \label{intdensc}
N(E_{0}) =\int_{0}^{E_{0}} \Omega(E) =
 2 (\tanh^{-1} \Omega(E_{0}) -\Omega(E_{0})).
\end{equation}
Inverting this relationship (cf. (\ref{brill})), we have
\begin{equation} \label{invtanhcomplex}
\Omega(E_{0}) = \tanh{{N(E_{0}) +2 \Omega(E_{0}) \over 2}},
\end{equation}
which for large $E_{0}$, yields
\begin{equation} \label{invtanhcomplexapprox}
\Omega(E_{0}) \approx \tanh{{N(E_{0}) + 2 \over 2}}.
\end{equation}

The Gibbs distributions
having $\Omega(E)$ as their structure function are
\begin{equation} \label{4}
 f(E; \beta) = { e^{- \beta E} \over  Z( \beta)} \Omega(E) \qquad
(0 \leq \beta \leq \infty),
\end{equation}
where $\beta$ is the {\it effective polarization}
 temperature parameter \cite{bross} and
\begin{equation} \label{5}
 Z(\beta) = { \sqrt {\pi} \Gamma (\beta) \over  2 \Gamma(3/2 + \beta)}
\end{equation}
serves as the partition function.
In general, the zeros of a partition function are of special interest
\cite{lee}. In this regard, let us
 note that the numerator of $Z(\beta)$ can
not assume the value zero,
 since the gamma function has no zeros in the complex plane,
although either its real or imaginary parts can vanish \cite{boh}.
However, the denominator of $Z(\beta)$ assumes infinite magnitude at the
[isolated] points
$\beta = -3/2,-5/2,-7/2,\dots$, since $\Gamma(x)$ has simple poles
at $x=-n,\quad n=0,1,2,\dots$ with residues $(-1)^{n}/n!$

We can compute the average or expected value of $E$, as
\begin{equation} \label{ave}
\langle E \rangle = - {\partial \over \partial \beta} \log Z(\beta)
        =\psi(3/2+\beta) -\psi(\beta)
        = \sum_{1}^{\infty} {(3/2)_{n} \over n (3/2+\beta)_{n}},
\end{equation}
where the digamma function, $\psi(z)= {d \over dz}{\log \Gamma(z)}
         = {\Gamma'(z) \over \Gamma(z)}$. (The last equality in (\ref{ave}),
 employing
the Pochhammer symbol $(\lambda)_{n}$ is taken from \cite{als}, where it
is also shown that this series can be written in
terms of a generalized hypergeometric
  ${}_{3}F_{2}$ series.)
For small values of $\beta$, we have
\begin{equation} \label{approx}
\beta \approx {1 \over \langle E \rangle  -2 - \log 2}.
\end{equation}
For large values of $\beta$,
\begin{equation} \label{bigapprox}
\beta \approx {3 \over 2 \langle E \rangle}.
\end{equation}
This can be understood from the
 fact that the digamma function satisfies the functional equation,
\begin{equation} \label{psi}
\psi(1+\beta)-\psi(\beta) = {1 \over \beta},
\end{equation}
or, more formally, from the asymptotic ($\beta \rightarrow
\infty$) expansion of $\langle E \rangle = \psi(3/2+\beta)-\psi(\beta)$,
\begin{equation} \label{asymptmaple}
{3 \over 2 \beta} -{3 \over 8 \beta^2} +{1 \over 4 \beta^3}
-{9 \over 64 \beta^4} +{1 \over 16 \beta^5} +O({1 \over \beta^6}).
\end{equation}
(It is highly interesting to note that one obtains the maximum-likelihood
estimator $\hat{\beta} = 3 / {2 \langle E \rangle}$ for an ideal monatomic
gas, for which the logarithm of the partition function is
$ - {3 \over 2} \log \beta$ \cite[p. 468]{lav1}.)
The variance of $E$ --- that is,
 $\langle (E - \langle E \rangle)^{2} \rangle$ --- is given by
\begin{equation} \label{var}
\mbox{var}(E) = {\partial^2 \over \partial \beta^2} \log Z(\beta)
= \psi'(\beta) - \psi'(3/2+\beta) {\approx {3 \over 2 \beta^{2}}}
\qquad (\beta \rightarrow \infty).
\end{equation}

Interchanging the roles of $E$ and $\beta$ \cite{mand,lav1,lav2,tik},
we obtain a (modal, as opposed to maximum likelihood \cite{mand}) estimate
 of the effective polarization temperature parameter $\beta$,
\begin{equation} \label{betmean}
  {\partial \over \partial E} \log \Omega(E)
= {1 \over 2 (e^E-1)}.
\end{equation}
Also, we have that
\begin{equation} \label{betvar}
 - {\partial^2 \over \partial E^2} \log \Omega(E)
 = {e^E \over 2 (e^E -1)^2}.
\end{equation}
\subsubsection{Approximate duals of the Gibbs canonical distributions}
\label{secdual}
The square root of (\ref{var}) is the indicated (minimally
 informative Bayesian)
 prior
over the parameter $\beta$ \cite{lav1}.
 It is most interesting to note that as $\beta \rightarrow \infty$, this is
 proportional to $1/\beta$.
The prior $1/\beta$ has arisen in several thermodynamic models and 
``is none other than Jeffreys' improper prior for $\beta$
 based on the invariance
property that the prior be invariant with respect to powers of $\beta$''
\cite{lav1} or ``the result obtained by Jeffreys for a scale parameter''
\cite{tik}.
The dual distribution (now regarding $E$ as the parameter to be
fixed, rather than $\beta$)
can then be written {\it approximately}
  (due to the use of (\ref{bigapprox})) as \cite{tik}
\begin{equation} \label{dual}
\tilde{f} (\beta;\langle E \rangle)
 \approx {3 e^{-\beta E} \sqrt {\mbox{var}(E)}
Z({3 \over 2 \langle E \rangle}) \over  2 Z(\beta)}.
\end{equation}
As an example of this (approximate)
 duality, we have set $\langle E \rangle=16.3$ in (\ref{dual}) and
divided the result by .984296 to obtain a probability distribution
over $\beta  \in  [0,\infty]$.
The expected value of $\beta$ was, then, found to be
 .0636579. Substituting this value
into (\ref{ave}), we obtain  an  $\langle E \rangle$ of 16.2805 --- which
is near to 16.3.
(For an initial choice of $\langle E \rangle < 16.3$, we would expect
the duality to be more exact in nature.)
Now, in general,
\begin{equation} \label{aveb}
{\langle \beta \rangle} \approx {{\partial \over \partial \langle E \rangle}
 \log Z({3 \over 2 \langle E \rangle})}
= {3 \over 2 {\langle E \rangle}^2}
 (\psi({3 \over 2} + {3 \over 2 \langle E \rangle})
- \psi({3 \over 2 \langle E \rangle}))
\approx {3 \over 2 \langle E \rangle}
\end{equation}
and
\begin{displaymath}
{\mbox{var}(\beta)} \approx {-{\partial^{2} \over \partial {\langle
 E \rangle}^{2}}
 \log Z({3 \over 2 \langle E \rangle})}
\end{displaymath}
\begin{equation}   \label{varb}
  = {3 \over 4 {\langle E \rangle}^{4}}
 (4 \langle E \rangle (\psi({3 \over 2} +{3 \over 2 \langle E \rangle})
-\psi({3 \over 2 \langle E \rangle})) + 3 (\psi'({3 \over 2} +{3 \over
2 \langle E \rangle}) - \psi'({3 \over 2 \langle E \rangle})))
\approx {3 \over 2 {\langle E \rangle}^{2}} \qquad (\langle E \rangle
\rightarrow 0).
\end{equation}
(The variance of $\beta$ for an ideal gas of $N$ particles is
${3 N \over 2 {\langle E \rangle}^2}$ \cite[p. 207]{lav2}.)
The square root of var($\beta$), which is approximately proportional to
${1 \over \langle E \rangle}$
(again, in conformity to Jeffreys' rule for a scale
parameter), then, serves as the prior over
 $\langle E \rangle$ \cite[eq. 21]{lav1}.
Following \cite[eq. 27a]{lav1}, we can alternatively
 attempt to express the prior over
$\langle E \rangle$ as the product of $\Omega(\langle E \rangle)$
and the reciprocal of $Z({3 \over 2 \langle E \rangle})$.
We have plotted this result along with ${1 \over \langle E \rangle}$ and both
curves display quite similar monotonically decreasing behavior.
\subsubsection{Thermodynamic interpretation of information-theoretic results
 of Krattenthaler and Slater}
We have been led to advance the model (\ref{3})-(\ref{5}) on the basis of
results reported in \cite{kratt}
and subsequent related analyses.
In \cite{kratt},
 a one-parameter family (denoted $q(u)$) of probability distributions over
the three-dimensional convex set (the ``Bloch sphere'' \cite{braun2},
 that is, the unit
ball in three-space) of two-level
quantum systems (\ref{1}) was studied. It took the form
\begin{equation} \label{6}
q(u) = { \Gamma(5/2-u) r^2 \sin \theta \over \pi^{3/2} \Gamma(1-u) (1-r^2)^u}  \qquad ( -\infty < u < 1).
\end{equation}
We note that the Gibbs distributions (\ref{4}) can be obtained
from (\ref{6}) through the pair of changes-of-variables
$u=1- \beta$ and $r= \sqrt {1-e^{-E}} = \Omega(E)$ (which is
 the positive branch
of (\ref{invert})),
 accompanied by integration over
the two angular coordinates ($0  \leq \theta < \pi,0 \leq \phi < 2 \pi$).
(We will in sec.~\ref{secquat} use this pair of transformations
 also in the quaternionic, real and
classical counterparts of the standard complex analysis.)
In Cartesian coordinates, $q(u)$ takes the form
\begin{equation} \label{cart}
{\Gamma(5/2-u) \over \pi^{3/2} \Gamma(1-u) (1-x^2-y^2-z^2)^u}
\end{equation}
Making the substitutions $X=x^2,Y=y^2$ and $Z=z^2$, this becomes a
Dirichlet distribution \cite{ferguson},
\begin{equation} \label{dirich}
{\Gamma(5/2-u) \over \pi^{3/2} \Gamma(1-u) \sqrt{X} \sqrt{Y} \sqrt{Z}
(1-X-Y-Z)^u},
\end{equation}
over the three-dimensional probability simplex spanned by the possible
values of $X,Y$ and $Z$.
(``In Bayesian statistics the Dirichlet distribution is known as the
{\it conjugate distribution} of the multinomial distribution...In this sense
the discrete coherent states are the quantum conjugate states of the
Bose-Einstein symmetric number states'' \cite[Remark 2.3.3]{bach}.)

The initial motivation for studying the family (\ref{6}) in \cite{kratt}
 was that it contained as
a specific member --- $q(.5)$ --- a
 probability distribution (the ``quantum Jeffreys prior'') proportional to the
volume element of the Bures metric \cite{braun} on the Bloch sphere.
(It now appears that
 all those values of $u$ lying between .5 and 1.5, or equivalently
$\beta \in$ [-.5,.5], correspond to such ``monotone''
 metrics \cite{petzsud,petz3}.
The desirability of monotone metrics stems from the finding that an
``infinitesimal statistical distance has to be monotone under stochastic
mappings'' \cite[p. 486]{PETZ}.)
We were interested in \cite{kratt} (cf. \cite{slatjpa}) in the possibility
(in analogy to certain classical/nonquantum results of
Clarke and Barron \cite{clarke}) that the distribution $q(.5)$
might possess certain information-theoretic
 {\it minimax} and {\it maximin} properties {\it vis-\`{a}-vis}
all other probability distributions over the Bloch sphere. 
However,  considerations of analytical tractability
 led us to examine only those
distributions in the family $q(u)$.

In \cite{kratt}, the probability distributions $q(u)$ were used to
average the $n$-fold tensor products $(\stackrel{n}{\otimes} \rho)$ over
 the Bloch
sphere, thereby obtaining a one-parameter family
of $2^n \times 2^n$ averaged matrices,
$\zeta_{n} (u)$. The eigenvalues of $\zeta_{n} (u)$ were
found to be
\begin{equation} \label{7}
\lambda_{n,d} = {1 \over 2^n} {\Gamma(5/2-u) \Gamma(2+n-d-u) \Gamma(1+d-u)
\over {\Gamma(5/2 +n/2-u) \Gamma(2+n/2-u) \Gamma(1-u)}},
\qquad d = 0,1,\dots,\lfloor {\frac {n} {2}} \rfloor
\end{equation}
with respective multiplicities,
\begin{equation} \label{mult}
 m_{n,d} = { (n-2d+1)^2 \over (n+1)}   {n+1 \choose d}.
\end{equation}
(Parallel analyses have also been conducted using the real analogue,
$q_{real}(u)$, given by formula (\ref{qreal}),
 of the standard complex probability distribution $q(u)$.
However, it has proved more difficult, in this case, to give a formal
demonstration of the validity of the derived expressions for the
corresponding eigenvalues and eigenvectors.)

The subspace --- spanned by all those eigenvectors
associated with the same eigenvalue $\lambda_{n,d}$ --- corresponds
to those explicit spin states \cite[sec. 7.5.j]{bied} \cite{pauncz}
with $d$ spins either ``up'' or ``down''
(and the other $n-d$ spins, of course, the reverse).
The $2^n$-dimensional Hilbert space can be decomposed into
the direct sum of carrier spaces of irreducible representations
of $SU(2) \times S_{n}$. The multiplicities (\ref{mult}) are
 the dimensions of the corresponding irreps. 
The $d$-th space consists of the union of ${m_{n,d} \over (n-2 d+1)}$
copies of irreducible representations of SU(2), each of dimension
$(n-2 d+1)$ or, alternatively, of $(n-2 d+1)$ copies of irreps of
$S_{n}$, each of dimension ${m_{n,d} \over (n-2 d+1)}$.

In \cite{kratt}, an explicit formula, in terms of the eigenvalues,
(\ref{7}) was obtained for
 the relative entropy of $\stackrel{n}{\otimes} \rho$
with respect to $\zeta_{n}(u)$,
\begin{equation} \label{relent}
-n S(\rho) - \mbox{Tr} (\stackrel{n}{\otimes} \rho \cdot  \log \zeta_{n}(u)),
\end{equation}
where $S(\rho) = - \mbox{Tr} \rho \log \rho$ is the von Neumann
entropy of $\rho$.
Then, the asymptotics of this quantity (\ref{relent})
 was found as $n \rightarrow \infty$.
We can now reexpress this result (\cite[eq. 2.36]{kratt})
 in terms of the thermodynamic
variables ($\beta, E$), rather than $r$ and $u$, as
(using the positive branch of (\ref{invert}))
\begin{equation} \label{asympt}
{3 \over 2} \log n -{1 \over 2} -{3 \over 2} \log 2 + \beta E
+ {1 \over 2 \Omega(E)} \log [{1- \Omega (E) \over 1+ \Omega(E)}]
+\log \Gamma(\beta) - \log \Gamma( {3 \over 2} + \beta)+ O({1 \over n}).
\end{equation}
 Let us naively (that is, ignoring the error term immediately above)
 attempt to find an asymptotically stationary point
of the relative entropy (\ref{relent})
 by setting the derivatives of (\ref{asympt})
 with respect to $\beta$ and
$E$ to zero, and then numerically solving the resultant
pair of nonlinear simultaneous equations. The derivative with
respect to $\beta$  simply yields the condition (\ref{ave}) --- that is,
the likelihood equation --- that $E = \langle E \rangle$, while that
 with respect to $E$ can be expressed as
\begin{equation} \label{DE}
\beta={1  \over 4 \Omega(E)^2} (2 + {1 \over e^E \Omega(E)} \log [{1-\Omega(E)
\over 1 + \Omega(E)}]).
\end{equation}
We obtained a stationary value of the truncated asymptotics
(\ref{asympt}) at the point
($\beta =.457407, E=2.58527$). It appears possible that this point serves
as an asymptotic minimax for the relative entropy. Though additional
numerical evidence appears consistent with such a proposition, a formal
demonstration remains to be given.
In \cite{kratt}, the asymptotics of the {\it maximin} --- with
respect to the one-parameter family --- of the
relative entropy was formally found.
(This involved integrating out the radial coordinate $r$. In this
analysis, it was possible to show that it was harmless
to ignore the error term in (\ref{asympt}).) It corresponded
 to the value
$\beta = .468733$. This  was obtained as the solution of an equation
\cite[eq. (3.10)]{kratt},
here expressible as
\begin{equation} \label{maximin}
2 \beta^3 \mbox{var}(E) = 1.
\end{equation}
(Use of $\mbox{var}(E) \approx {3 \over 2 \beta^{2}}$ from (\ref{var})
in this equation
would give a solution of $\beta = {1 \over 3}$.)
It is interesting to note that both the [rather
proximate] values of $\beta$ of .457407
and .468733 lie in the range [-.5,.5] associated with monotone
metrics \cite{petzsud}. In their classical/nonquantum analysis,
Clarke and Barron \cite{clarke} found that both the asymptotic
minimax and maximin were given by the very same probability distribution ---
the Jeffreys prior, which is based on the {\it unique} monotone metric
in that classical domain of study.

Let us note --- in view of the forms of expression of (\ref{asympt}) and
(\ref{DE}) --- that in \cite[eq. (2.4)]{bross}
 (cf. \cite{christ}) the expression
\begin{equation} \label{bros}
{1 \over 2} \log [ {1+\Omega(E) \over 1-\Omega(E)}]
= {1 \over 2} \log[{1+r \over 1-r}]
\approx r +{r^3 \over 3} +{r^5 \over 5} + \ldots
\end{equation} 
was taken to be the reciprocal ($\tau^{-1}$)
 of ``an effective polarization temperature ($\tau$)
[which] should not be confused with the radiance temperature obtained
using Planck's spectral law''.
(The equality in (\ref{bros}) will {\it not} hold in the quaternionic
and real cases discussed below (sec.~\ref{secquat}), since then the structure
function will not simply equal
 the radial coordinate.) Inverting equation (2.4) of \cite{bross}, we obtain
\begin{equation} \label{brosseq}
r ={e^{2 / \tau} - 1 \over e^{2/ \tau} +1} = \tanh{{1 \over \tau}}
\end{equation}
(cf. formulas (1.28) and (1.36) of \cite{balian}).
It will be of substantial interest
(setting $\tau=\beta$) to compare (\ref{brosseq}) with
the relations we obtain --- (\ref{expO}), (\ref{quatpolar}) and
(\ref{realexp}) --- for the average polarization $\langle r \rangle$
 as a function of $\beta$.
This will be done in Fig.~\ref{fig_fog1}.

\subsubsection{Dependence of the
 average polarization upon the parameter $\beta$}
 The expected value of the
structure function $\Omega(E)$, that is (\ref{3}) --- equal to the radial
coordinate ($r$) --- with respect to the family
of Gibbs distributions (\ref{4}) is exactly computable. We have that
\begin{equation} \label{expO}
\langle r \rangle =
\langle {\Omega(E)} \rangle =
{1 \over \beta (1+ \beta) Z(\beta)} =
{2 \Gamma(3/2+ \beta) \over \beta (1 + \beta) \sqrt {\pi} \Gamma(\beta)}
 = {2 \Gamma(3/2 + \beta) \over \sqrt {\pi} \Gamma(2 + \beta)}
\end{equation}
For small values of $\beta$ (cf. (\ref{approx})),
\begin{equation} \label{approx2}
\beta \approx {1 - \langle r \rangle \over 2 \log 2 - 1}.
\end{equation}
Asymptotically, $\beta \rightarrow \infty$,
\begin{equation} \label{meanfield}
\langle r \rangle = {1 \over \sqrt{\pi}} ({2 \over \beta^{1/2}}
-{5 \over 4 \beta^{3/2}} +{73 \over 64 \beta^{5/2}} -{575 \over 512
\beta^{7/2}}) +O({1 \over \beta^{9/2}}).
\end{equation}
We have found a
 series  of curves for integral $n$ that converges
(as C. Krattenthaler has formally demonstrated)
to (\ref{expO}) as $n \rightarrow \infty$ (cf. \cite{farhi}).
The curves are obtained by summing the weighted
terms $(n - 2 d)/n$ over the
 $1+ \lfloor n/2 \rfloor$ subspaces (indexed by $d$) of explicit spin states,
 using as the weights, the probabilities assigned to
the subspaces, that is,  the product of the
eigenvalue ($\lambda_{n,d}$) for the $d$-th subspace (\ref{7})
 and its associated multiplicity ($m_{n,d}$)
(\ref{mult}). Thus, we consider
\begin{equation} \label{spin}
\sum_{d=0}^{\lfloor n/2 \rfloor} ({n-2 d \over n}) m_{n,d} \lambda_{n,d}.
\end{equation}
  This gives the average or expected polarization (since $n-2d$
is the net spin for $n-d$ majority spins assigned $+1$ and $d$ 
minority spins assigned $-1$), ranging between
0 and 1. (There have been recent studies \cite{chak,man}
 of the temperature dependence of
the average
 spin polarization in a system  --- a quantum Hall ferromagnet --- in
 which the spins are {\it not} [as assumed
here]
{\it noninteracting} (cf. \cite{riddle}).)
Thus, similarly to the parameter $\tau$ in \cite{bross}, the value $\beta=0$
 corresponds to complete polarization
(total domination by the majority spin)
and $\beta= \infty$ to
the case in which there is no predominance of spin in
any direction.

The convergence of the sum (\ref{spin}) to the mean value (\ref{expO})
has been  essentially
 demonstrated by Krattenthaler by, first, rewriting (\ref{spin})
as the difference
\begin{equation} \label{diff}
\sum_{d=0}^{\lfloor n/2 \rfloor} ({n-2 d +1 \over n}) m_{n,d} \lambda_{n,d}
-\sum_{d=0}^{\lfloor n/2 \rfloor} ({1 \over n}) m_{n,d} \lambda_{n,d}.
\end{equation}
(His proof --- applicable to any probability distribution over the Bloch
sphere which is spherically-symmetric --- is
 conducted in terms of the spherical coordinates used in
\cite{kratt}, but it easily carries over to the presentation here in
terms of the thermodynamic variables, $\beta$ and $E$.)
He shows --- employing Theorem 15 of \cite{kratt} to reexpress
$\lambda_{n,d}$, then splitting the resulting expression into four terms,
in the manner of (2.41)-(2.46) of \cite{kratt} and applying the
binomial theorem --- that (\ref{diff}) and, hence, (\ref{spin})
can be rewritten as the sum of
the expected value of $q(u)$ (that is, ${2 \Gamma (5/2-u) \over
\sqrt{\pi} \Gamma(3/2-u)}$) and a term that converges to zero
as $n \rightarrow \infty$.
\subsubsection{Relations between average polarization
(\ref{expO}) and
 hyperbolic tangent Brillouin (\ref{brill}) functions}
\begin{figure}
\centerline{\psfig{figure=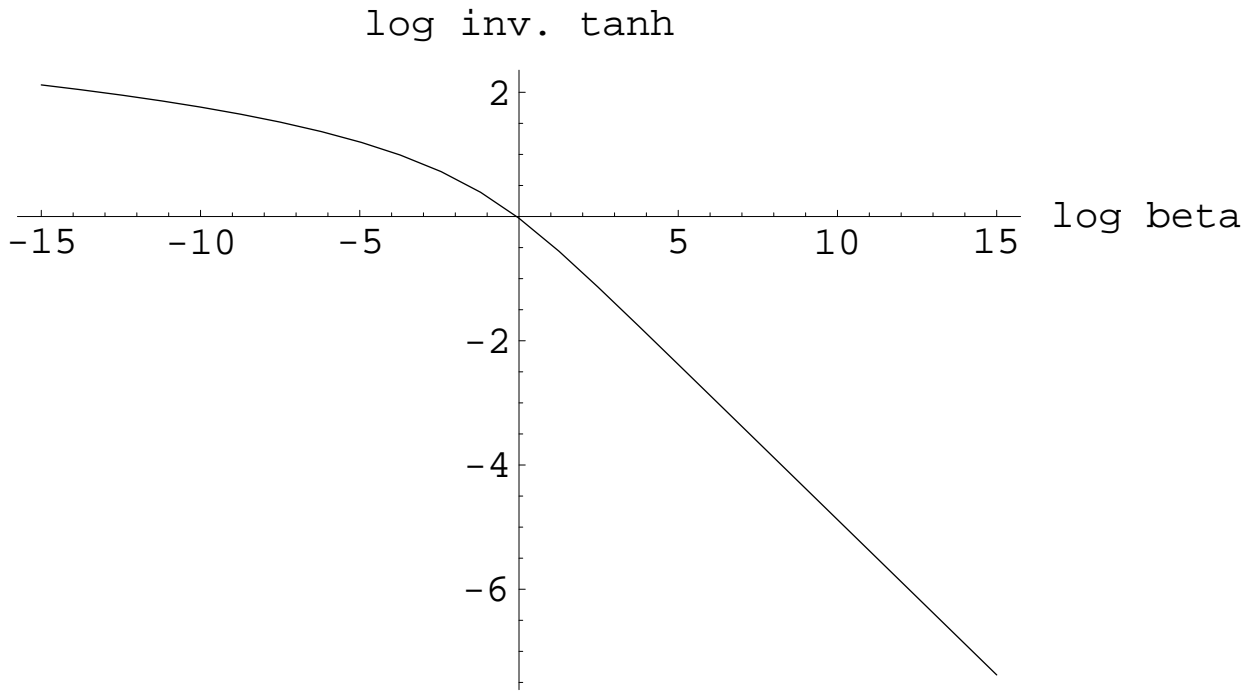}}
\vspace{-2in}
\caption{The Logarithm of the Inverse Hyperbolic Tangent of the
Average Polarization (\ref{expO}) for the Standard Complex Model of
sec.~\ref{sc}
{\it vs}. the Logarithm of the Effective Polarization Temperature 
Parameter $\beta$.
For large $\beta$, the curve is approximately linear with a slope of
$-{1 \over 2}$.}
\label{plot6}
\end{figure}

Let us, in the obvious task of trying to relate the results here to the
conventionally employed hyperbolic tangent
Brillouin function (\ref{brill}), equate the reduced magnetization
\cite{tusz} given by ${M \over M_{0}} = \tanh{{\mu H \over k T}}$ to
 the average
polarization
 $\langle r \rangle$ obtained from formula (\ref{expO}).
To do so, we set the reduced temperature ($x \equiv {\mu H \over k T})$
equal
to
$\tanh^{-1} \langle r \rangle$. Asymptotically then, as
  $\beta \rightarrow \infty$, this gives
\begin{equation} \label{xbeta}
x ={2 \over \sqrt{\pi} \sqrt{\beta}}+{32 -15 \pi \over 12 {\pi}^{3/2}
{\beta}^{3/2}} +O({1 \over {\beta}^{2}}).
\end{equation}
Consequently, for large values of
 $\beta$, we have the log-linear approximation,
\begin{equation} \label{xbetalog}
\log{x} \approx \log{2} -{1 \over 2} \log{\pi} -{1 \over 2} \log{\beta}
= .120782 - {1 \over 2} \log{\beta}.
\end{equation}
In Fig.~\ref{plot6}, we  plot
$\log{\tanh^{-1} \langle r \rangle}$ {\it vs}. the logarithm of
the effective polarization temperature $\beta$.
For large $\beta$, the curve is approximately linear, having a
slope of $-{1 \over 2}$.

 To further study the relationship of the results in this study to the
standard treatment of paramagnetism (see also the formulas we obtain
for the integrated density of states --- (\ref{invtanhcomplex}),
(\ref{approxquatint}), (\ref{classinv})), in which the hyperbolic
tangent Brillouin function (\ref{brill})
 is employed \cite{tusz,schlogl,gold,bal2},
 it might prove useful
to employ
the relationships between the hyperbolic functions and the gamma functions
in the complex plane \cite[eqs.~6.1.23-6.1.32]{abram}.
Pursuing this line, we have found that
\begin{equation} \label{complex}
\langle r \rangle {2 \sqrt{\pi}
 (\beta+1) \Gamma(3/2-\beta) \over
 (4 {\beta}^{2} -1) \Gamma(-\beta)} =i \tanh {\pi \beta \over i}
=\tan \beta \pi.
\end{equation}
Error bounds for the asymptotic expansion of the ratio of two gamma functions
with complex argument (cf. (\ref{5}) and (\ref{expO}))
have been given in \cite{frenzen} (extending earlier work \cite{frenzen2} on the ratio
of two gamma functions with {\it real} argument), using generalized Bernoulli
polynomials (cf. \cite{paris}).
\subsection{Cases other than the standard complex} \label{secquat}
Although quantum mechanics is usually
treated as a theory over
the algebraic field of complex numbers, it is also possible and
 of considerable interest
to study versions over the real and quaternionic fields,
as well \cite{adler,slaterjmp}.
The possibility of testing these different forms against one another
has been studied \cite{fivel,davies,brumby}.
 It appears that the results here --- giving {\it distinct} Gibbs
distributions in the three quantum mechanical scenarios  --- may present
another avenue to addressing such fundamental issues.
\subsubsection{Quaternionic case}
For the quaternions, the analogue of the probability distribution
$q(u)$ given in (\ref{6}) takes the form (cf. \cite[eq. 26]{slaterjmp}),
\begin{equation} \label{quatqu}
q_{quat}(u) = {\Gamma(7/2-u) r^4 {\sin}^3 \theta_{1} {\sin}^2 \theta_{2}
\sin \theta_{3} \over \pi^{5/2} \Gamma(1-u) (1-r^{2})^{u}} \qquad
(-\infty <u <1),
\end{equation}
being defined over the {\it five}-dimensional unit ball or ``quaternionic
Bloch sphere.''
Proceeding with the same transformations as in the standard
complex case --- that is,
$u=1-\beta$ and $r=\sqrt{1-e^{-E}}$ --- and integrating over the four
angular coordinates, we arrive at a family of Gibbs
distributions (\ref{4}) having as their structure function or density
of states (cf. (\ref{3})),
\begin{equation} \label{quatstruct}
\Omega_{quat}(E) = \Omega_{complex}^{3}(E) = (1-e^{-E})^{3 \over 2}
\end{equation}
(which is approximately $E^{3/2}$ for small $E$),
where $\Omega_{complex}(E) \equiv \Omega(E) = \sqrt{1-e^{-E}}$, from
(\ref{3}).
The integrated density of states (cf. (\ref{intdensc})) is, then,
\begin{equation} \label{intdenquat}
N_{quat}(E_{0}) = \int_{0}^{E_{0}} \Omega_{quat}(E) =
2 (\tanh^{-1} \Omega_{complex}(E_{0}) +{\Omega_{complex}(E_{0}) (1-4 e^{E_{0}})
\over 3 e^{E_{0}}}),
\end{equation}
Also, we have the difference,
\begin{equation} \label{diffden}
N_{complex}(E_{0}) -N_{quat}(E_{0}) = {2 \Omega_{complex}(E_{0}) (e^{E_{0}}-1)
\over 3 e^{E_{0}}}.
\end{equation}
As $E_{0} \rightarrow \infty$, it approaches from below, the limiting value of
 ${2 \over 3}$.
For large $E_{0}$ (cf. (\ref{brill}), (\ref{invtanhcomplexapprox})),
\begin{equation} \label{approxquatint}
 \Omega_{complex}(E_{0}) \approx \tanh{{3 N_{quat}(E_{0}) + 8 \over 6}}.
\end{equation}

The partition function (cf. (\ref{5})) in the quaternionic case is
\begin{equation} \label{quatpart}
Z_{quat}(\beta) = {3 \sqrt{\pi} \Gamma(\beta) \over 4 \Gamma(5/2+\beta)}.
\end{equation}
We now have (cf. (\ref{ave}))
\begin{equation} \label{quatave}
{\langle E_{quat} \rangle} = - {\partial \over \partial \beta}
 \log Z_{quat}(\beta)
=\psi(5/2+\beta) -\psi(\beta) \approx {5 \over 2 \beta}
\qquad (\beta \rightarrow \infty)
\end{equation}
and (cf. (\ref{var}))
\begin{equation} \label{quatvar}
\mbox{var}_{quat}(E) ={\partial^2 \over \partial {\beta}^{2}}
\log Z_{quat}(\beta) =
\psi'(\beta) - \psi'(5/2 + \beta) \approx {5 \over 2 {\beta}^{2}}
\qquad (\beta \rightarrow \infty).
\end{equation}
It, thus, appears that the quaternionic results relating to approximate
duality can, in essence, be
obtained from the (standard
 complex) ones given above (sec.~\ref{secdual}),
 simply through the substitution
of ${5 \over 2}$ for ${3 \over 2}$, as appropriate.
We have conducted a test of approximate duality
 similar to that reported (for the standard complex case) immediately after
(\ref{dual}). We have set ${\langle E \rangle}_{quat} = 16.3$ in the evident
quaternionic analogue of (\ref{dual}), and divided the result by .902062 to
obtain a probability distribution over $\beta \in [0,\infty]$.
The expected value of $\beta$ was, then, found to be .0664174.
Substituting this value into (\ref{quatave}), we obtain an estimate of
${\langle E \rangle}_{quat}$ equal to 16.2645 $\approx$ 16.3.

For the average
quaternionic polarization, we have
\begin{equation} \label{quatpolar}
{\langle r_{quat} \rangle} =
\langle {\Omega_{quat}(E)}^{1/ 3} \rangle = {8 \Gamma(5/2+\beta) \over
3 \beta (1 +\beta) (2 + \beta) (3 +\beta) \sqrt{\pi} \Gamma(\beta)}
= {8 \Gamma(5/2+\beta) \over 3 \sqrt{\pi} \Gamma(3 +\beta)}.
\end{equation}
(It would appear problematical, however, to develop a quaternionic
counterpart to the analysis of Krattenthaler
and Slater \cite{kratt},
as there are difficulties in defining an appropriate extension of
 the tensor product
\cite{adler}.)
\subsubsection{Real case}
We can also, proceeding similarly, take
\begin{equation} \label{qreal}
q_{real}(u)= {2 \Gamma(2-u) r \over \Gamma(1-u) (1-r^2)^u} =
 {(1-u) r \over \pi (1-r^2)^{u}}
\qquad (-\infty < u <  1),
\end{equation}
as a probability distribution over the unit (two-dimensional) disk.
Integrating over the {\it single} angular coordinate and using the
transformations,  $u=1- \beta$ and $r =\sqrt{1-e^{-E}}$,
we
transform (\ref{qreal}) to a Gibbs
distribution, having a structure function $\Omega_{real}(E) \equiv 1$
and a partition function,
\begin{equation} \label{partreal}
 Z_{real}(\beta) ={\Gamma(\beta)
\over \Gamma(1+\beta)} = {1 \over \beta}.
\end{equation}
(The integrated density of states is, of course, then, simply
$N_{real}(E_{0}) =E_{0}$.)
Now (cf. (\ref{psi})),
\begin{equation}
\langle E_{real} \rangle = -{\partial \over \partial \beta} \log Z_{real}
(\beta) = \psi(1+\beta)-\psi(\beta) = {1 \over \beta}
\end{equation}
So, it appears that the ``fraction''' $1 = {2 \over 2}$
 plays the analogous role
in real quantum mechanics as ${3 \over 2}$ in the standard complex
version and ${5 \over 2}$ in the quaternionic form.
(These results appear consistent with the equipartition of energy
theorem \cite{beghian}
 in classical statistical mechanics, as two, three and five
are the corresponding degrees of freedom, that is, the number of variables
needed to fully specify a real, standard complex or quaternionic density
matrix.)

 The expected value of $r$ with respect to
(\ref{qreal})  (again using $u=1- \beta$)
 gives us the average polarization in the real case,
\begin{equation} \label{realexp}
{\langle r_{real} \rangle} =
{\sqrt{\pi} \Gamma(1+\beta) \over 2 \Gamma( 3/2+\beta)}.
\end{equation}
The three forms of  average polarization --- (\ref{expO}),
(\ref{quatpolar}) and (\ref{realexp}) --- are particular instances of the
formula,
\begin{equation} \label{genform}
{\Gamma(1+m/2) \Gamma(1/2+ \beta +m/2) \over \Gamma(1+\beta +m/2)
\Gamma({1 + m \over 2})},
\end{equation}
where the quaternionic case corresponds to $m=4$, the standard complex to
$m=2$ and the real to $m=1$.
\subsubsection{Classical/nonquantum case}
The case $m=0$ is, essentially classical/nonquantum in nature,
corresponding to the family of probability distributions,
\begin{equation} \label{beta}
q_{class}(u) ={2 \Gamma(3/2-u) \over \sqrt{\pi} \Gamma(1-u) (1-r^2)^{u}}
\qquad (-\infty < u <1).
\end{equation}
(If one makes the transformation $r=\sqrt{R}$, this becomes
 (cf. (\ref{dirich})) a
family of beta distributions. Then, the member corresponding to
 $u = {1 \over 2}$,
 is the (bimodal) arcsine (alternatively, cosine
\cite{D'Ariano}) distribution, corresponding to a process with a single
degree of freedom, to which Boltzmann's principle is
inapplicable \cite{lav6}. The arcsine distribution serves as the
noninformative Jeffreys' prior for the Bayesian inference of the parameter
of a binomial distribution \cite{bern,slat7}.)
Making the same substitutions as in the previous cases,
 that is $u=1- \beta$ and $r=\sqrt{1-e^{-E}}$, into
 (\ref{beta}),
we obtain
\begin{equation} \label{classstruct}
 \Omega_{class}(E)  = 
 {1 \over \Omega_{complex}(E)} = {1 \over \sqrt{1-e^{-E}}},
\end{equation}
\begin{equation} \label{classpart}
 Z_{class}(\beta) = {\sqrt{\pi} \Gamma(\beta) \over \Gamma(1/2+\beta)},
\end{equation}
and
\begin{equation} \label{avgclass}
\langle E_{class} \rangle = -{\partial \over \partial \beta}
\log Z_{class}(\beta) =\psi(1/2+\beta) -\psi(\beta)
\approx {1 \over 2 \beta}
\qquad (\beta \rightarrow \infty)
\end{equation}
(corresponding to the {\it single} degree of freedom in this case).
Also, setting $m=0$ in (\ref{genform}), we have
\begin{equation} \label{rclass}
\langle r_{class} \rangle = {\Gamma(1/2+\beta) \over \sqrt{\pi}
\Gamma(1+ \beta)}.
\end{equation}
For the integrated density of states, we find
$N_{class}(E_{0}) = 2 \tanh^{-1} \Omega_{complex}(E_{0})$,
so 
\begin{equation} \label{classinv}
\Omega_{complex}(E_{0}) =\tanh{{N_{class}(E_{0}) \over 2}}.
\end{equation}
\subsection{Comparison of results for the four different cases}
The average polarization indices for the four models considered above
 all equal 1 for $\beta=0$ and monotonically decrease to
 0 for $\beta = \infty$.
However, for $\beta$ strictly between 0 and $\infty$, the
quaternionic value is
always greater than the standard complex one, which, in turn, is always
greater than the real index
 (which itself dominates the classical one). (In this regard, it may be
helpful to note that
 the ratio of (\ref{quatpolar}) to
(\ref{expO}) simplifies to ${2 (3+2 \beta) \over 3 ( 2 + \beta)}$.)

\begin{figure}
\centerline{\psfig{figure=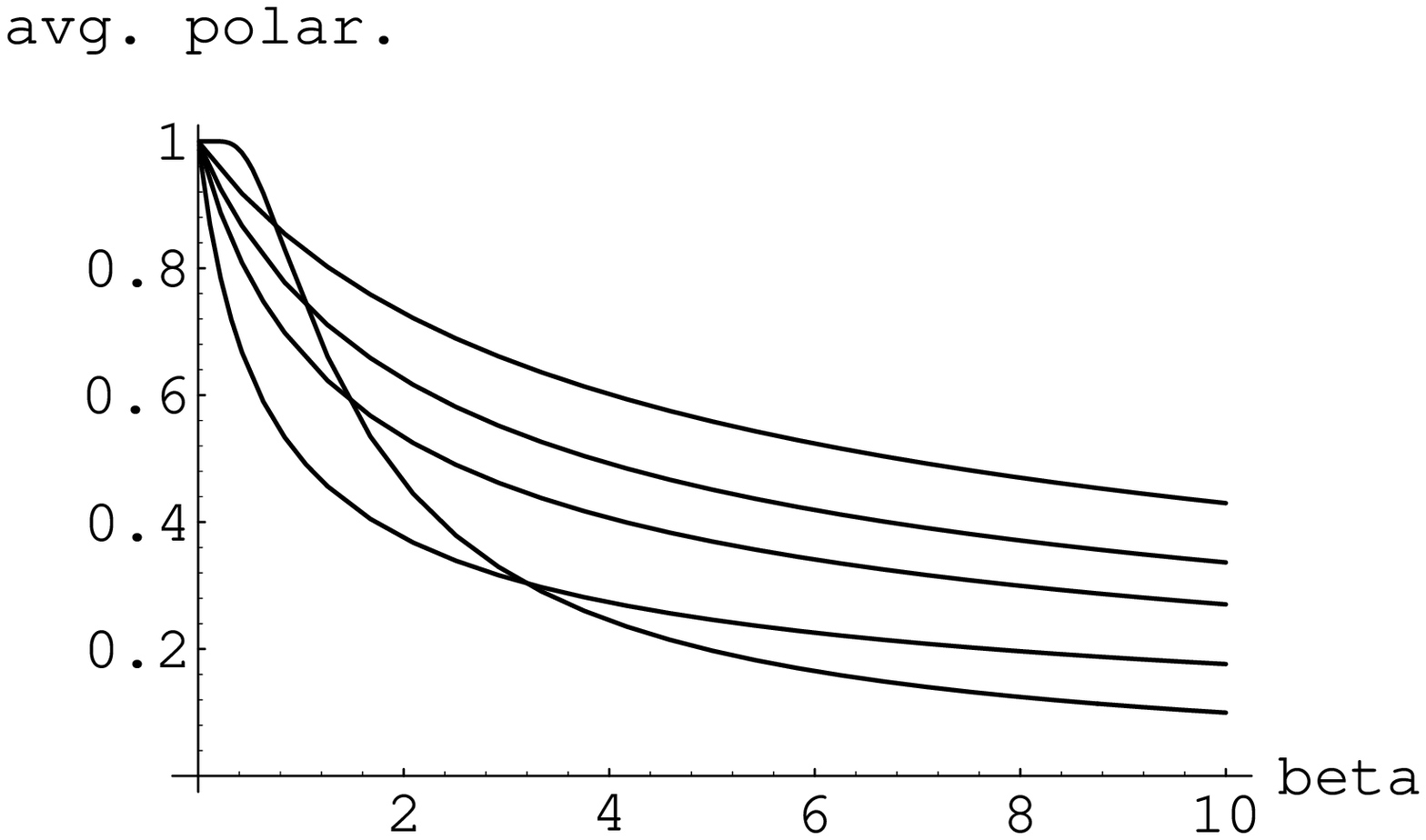}}
\vspace{-2in}
\caption[Average Polarization as a Function of
 $\beta$]{Average Polarization
 as a Function of the {\it Effective Polarization} Temperature
$\beta$.
 The quaternionic curve (\ref{quatpolar}) dominates the
standard complex curve (\ref{expO}), which, in turn, dominates the
real curve (\ref{realexp}). This dominates the classical/nonquantum
 curve
(\ref{rclass})). The function of Brosseau and Bicout
(\ref{brosseq}), setting $\tau=\beta$,
making it equivalent to $\tanh{{1 \over \beta}}$,
 crosses the four other curves,
intersecting at $\beta$ = .76007, 1.04585, 1.46249 and 3.1857.}
\vskip .2cm
\label{fig_fog1}
\end{figure}

In Fig.~\ref{fig_fog1}, we plot the average
 polarization as a function of $\beta$
for the quaternionic, standard complex, real and classical cases ($m=4, 2, 1$
and 0, respectively).
We also display the Brosseau-Bicout
 relation (\ref{brosseq}), having set $\tau=\beta$.
This last curve crosses the other four. It predicts greater average
 polarization in the vicinity of $\tau = \beta = 0$ and less above a
certain threshold (which is smallest in the quaternionic case and
largest in the classical one).

\begin{figure}
\centerline{\psfig{figure=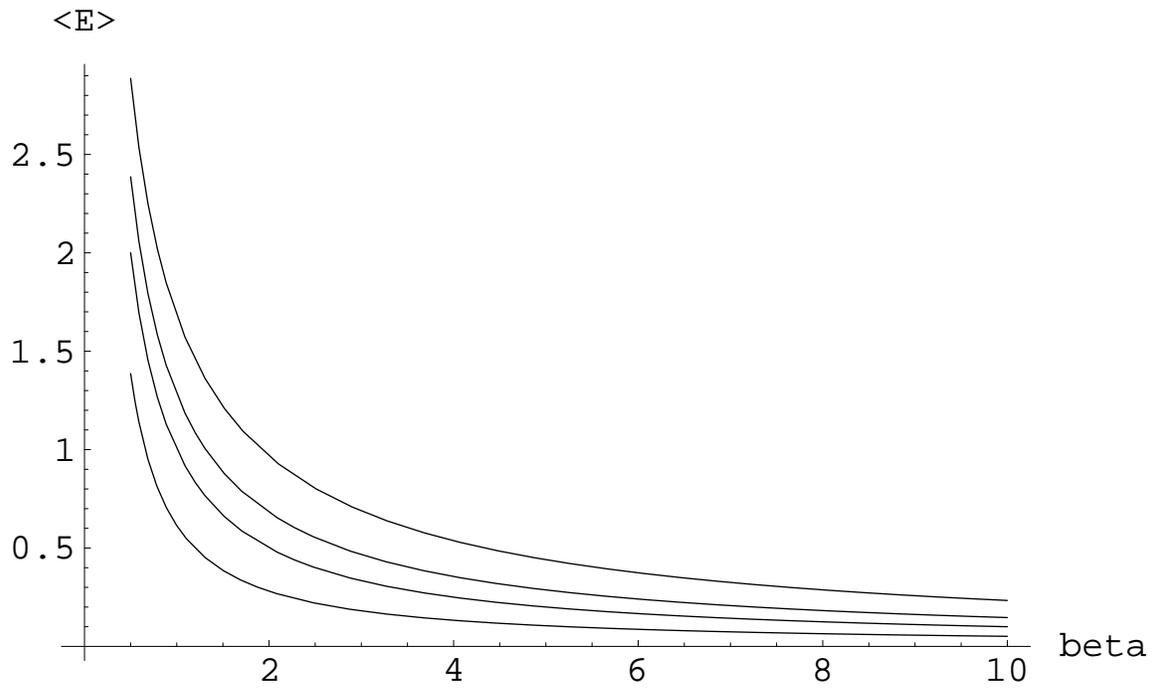,width=6in}}
\caption{Expected values $\langle E \rangle$ of the energy $E$.
The order of dominance is
 $\langle E_{quat} \rangle > \langle E_{complex}
\rangle > \langle E_{real} \rangle > \langle E_{class} \rangle$.}
\label{fig2}
\end{figure}

\begin{figure}
\centerline{\psfig{figure=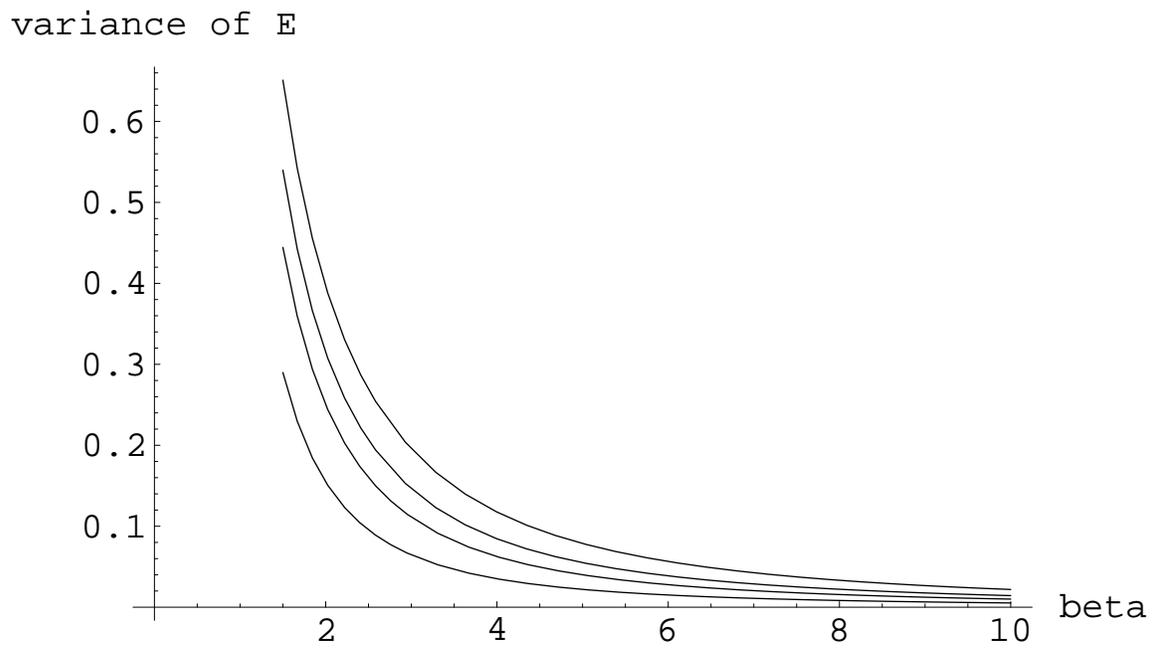,width=6in}}
\caption{Variances of the energy $E$. The order of dominance is
$\mbox{var}(E_{quat}) >\mbox{var}(E_{complex}) >\mbox{var}(E_{real})
> \mbox{var}(E_{class})$.}
\label{fig3}
\end{figure}

In Fig.~\ref{fig2}, we plot the expected value $\langle E \rangle$
of the energy $E$ for the quaternionic, standard complex, real
and classical cases, while in Fig.~\ref{fig3}, we show the
corresponding variances var($E$)..

\subsection{An alternative family of Gibbs  canonical distributions for the
standard complex case}
In the analyses reported above, we have relied upon certain
transformations of the probability distribution $q(u)$ (\ref{6}),
 defined over the
Bloch sphere of two-level systems, and its real (\ref{qreal}) and quaternionic 
  (\ref{quatqu})
analogues, to obtain the Gibbs distributions we have studied.
The distribution $q(u)$ has the attractive feature that for $u \in [.5,1)$
(and also for $u \in [1,1.5]$, although $q(u)$ is improper in this range),
it is proportional to the volume element of a {\it monotone} metric on the 
two-level systems. (The value $u=.5$ corresponds to the {\it minimal} [Bures]
monotone metric and $u=1.5$ to the {\it maximal} one.)

Nevertheless, there appear to be alternatives to $q(u)$ that are of certain
interest and might conceivably be physically meaningful. One such family
of probability distributions can be expressed as \cite[eq. (3.11)]{kratt}
\begin{equation} \label{KMB}
q_{KMB}(u) =
{(1-u) \Gamma(3/2-u) r \log [(1+r)/(1-r)] \sin \theta \over 2
\pi^{3/2}  \Gamma(1-u) (1-r^2)^{u}} \qquad (-\infty < u <1).
\end{equation}
For $u={1 \over 2}$, this is proportional to the volume element of
the Kubo-Mori/Bogoliubov (monotone) metric \cite{PETZ,petztoth,petzhas}.
``This metric is infinitesimally induced by the (nonsymmetric) relative
entropy functional or the von Neumann entropy of density matrices. Hence
its geometry expresses maximal uncertainty'' \cite{PETZ}.
The family (\ref{KMB}) can essentially be obtained from the family
$q(u)$ by replacing a factor of $r$ by $\log [{(1+r) \over (1-r)}]$
and renormalizing the result. For values of $r$ in the neighborhood
of 0, $q(u)$ and $q_{KMB}(u)$, for any fixed $u$, 
 are approximately proportional
(cf. (\ref{bros})).
(The KMB-metric is the extreme element [$\alpha = \pm 1$]
 of a one-parameter ($\alpha$) family --- distinct
from (\ref{KMB}) --- {\it all} the members of which correspond to monotone
Riemannian
metrics  induced by quantum alpha-entropies
 \cite{petzhas}. However, the corresponding volume elements for
 $-1 < \alpha <1$ do not
appear to be exactly
 normalizable over the Bloch sphere of two-level systems.)

\begin{figure}
\centerline{\psfig{figure=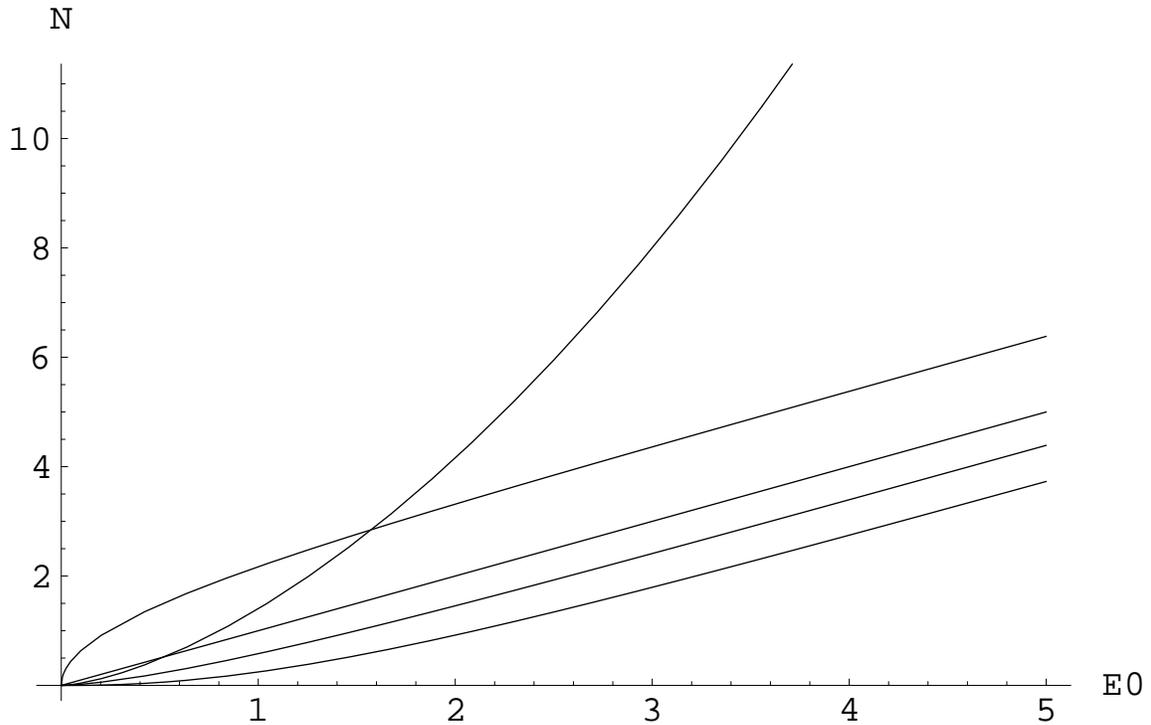,width=6.0in}}
\caption{Integrated Densities of States for Five Different Models.
 The most steeply-rising curve
corresponds to $N_{KMB}(E_{0})$, while the order of dominance for the other
four curves is classical $>$ real $>$ standard complex $>$ quaternionic.
The $KMB$-curve crosses the classical one at $E_{0}=1.57565$, the real linear
 one --- $N_{real}(E_{0}) = E_{0}$ --- at
.53341, the standard complex curve at .000111286, and the quaternionic
 curve at .0000405489. Asymptotically ($E_{0} \rightarrow \infty$), the
difference (\ref{diffden}), that is
$N_{complex}(E_{0}) - N_{quat}(E_{0})$, monotonically
 increases to its limit ${2 \over 3}$.}
\label{fig5}
\end{figure}

If we perform the transformations, $u=1-\beta$ and $r=\sqrt{1-e^{-E}}$,
 then (\ref{KMB}) becomes a family of Gibbs distributions (\ref{4}) with
 the partition function,
\begin{equation} \label{KMBpart}
 Z_{KMB}(\beta) = {Z_{class}(\beta) \over \beta} =
 { \sqrt{\pi} \Gamma(\beta) \over \beta \Gamma(1/2+ \beta)},
\end{equation}
and, quite interestingly,  the structure function
 (cf. (\ref{bros}),
\begin{displaymath}
\Omega_{KMB}(E) = \log [{(1+\sqrt{1-e^{-E}}) \over (1-\sqrt{1-e^{-E}})}
\end{displaymath}
\begin{equation} \label{KMBstruct}
= 2 (  \sqrt{1-e^{-E}} + {(1-e^{-E})^{3/2} \over 3} + \ldots)
=2 (\Omega_{complex}(E) +{\Omega_{quat}(E) \over 3} + \ldots).
\end{equation}
Using the positive branch of (\ref{invert}), $\Omega_{KMB}(E)$ is
 simply twice the reciprocal ($\tau^{-1}$) of the effective
polarization temperature ($\tau$), as defined by Brosseau and Bicout
\cite[eq. 2.4]{bross} (cf. (\ref{bros})).
The range of possible values of this structure function is the nonnegative
real axis $[0,\infty]$, in contrast
 to the structure functions considered above,
which only assume values on the unit interval [0,1]. We were able to
explicitly compute
the corresponding integrated density of states --- but the expression
MATHEMATICA 3.0 yielded was highly complicated.
In Fig.~\ref{fig5}, we have plotted the five integrated density
of states  functions obtained
here --- $N_{complex}(E_{0}), N_{quat}(E_{0}), N_{real}(E_{0})=E_{0},
N_{class}(E_{0})$ and $N_{KMB}(E_{0})$.
The first term of the asymptotic expansion ($\beta \rightarrow \infty$) of
$-{\partial \over \partial \beta} \log Z_{KMB}$ is
${3 \over {2 \beta}}$. This is the same as was found above (\ref{asymptmaple})
for the standard complex case and which coincides with the behavior of
the ideal monatomic gas.
The first term of the asymptotic expansion ($\beta \rightarrow \infty$)
of the square root of ${\partial^2 \over \partial {\beta}^{2}} \log Z_{KMB}$
is $ {3 \over 2 \beta^{2}}$, again as in the standard complex case
considered previously (\ref{var}).

\begin{figure}
\centerline{\psfig{figure=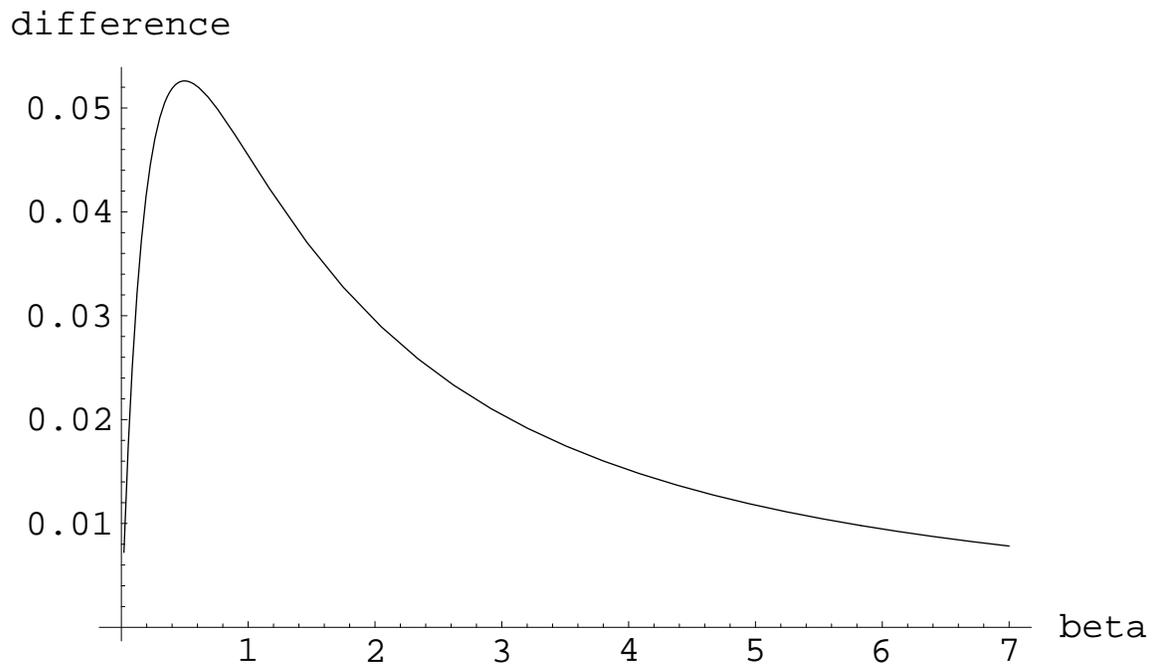,width=6in}}
\caption{The average polarization given by $\langle r_{KMB} \rangle$
minus the average polarization $\langle r \rangle$ --- given by (\ref{expO}),
 plotted
in Fig.~\ref{fig_fog1} --- for
 the first/primary standard complex model considered  here (sec.~\ref{sc}).}
\label{fig4}
\end{figure}

The expected value of $r$ with respect to (\ref{KMB}) --- giving us
the average polarization index --- can be expressed as
\begin{equation} \label{polKMB}
{\langle r_{KMB} \rangle} =
{2 \beta \Gamma(1/2 +\beta)  _{P}F_{Q}[\{{1 \over 2},1,2 \},
\{{3 \over 2},2 + \beta \}, 1]
\over \sqrt{\pi} \Gamma(2 +\beta)}.
\end{equation}
Due to its hypergeometric character, it is difficult to analytically
study (\ref{polKMB}).  A plot (Fig.~\ref{fig4}) of the difference between it
 and the alternative
(standard complex)
 average polarization index (\ref{expO}), shows  that the latter
index is
dominated by (\ref{polKMB}), with the greatest difference
($\approx .0526$) occurring in the vicinity of $\beta = .49825$.
(We have earlier indicated that for the distributions $q(u)$, given
in (\ref{6}), taking $u = 1 - \beta$, the range $\beta \in [-.5,.5]$ --- which
includes .49825 --- parameterizes a continuum of monotone metrics.)

\section{Concluding Remarks} \label{Conclusions}

The Brillouin functions of paramagnetism serve as the basis of the
mean-field theory of ferromagnetism \cite{tusz,schlogl,stanley}. It
 would, therefore, be of interest
to construct mean-field theories based on the alternative functions
that we have developed above. In this regard, following the usual line
of argument \cite[sec. 6.2]{stanley} \cite[sec. 9.2]{balescu},
 we have replaced $\beta$ by
 $\beta /(\lambda \langle r \rangle)$ in the small-$\beta$
 approximation
(\ref{approx2}) to the average polarization index (\ref{expO}) in
the standard complex case. ($\lambda$ corresponds to the molecular
field parameter.) Then, we find that there exists a nonzero critical
effective polarization temperature, 
\begin{equation} \label{critical}
\beta_{c} = {\lambda \over
4 (2 \log{2} - 1)} = 
.647175 \lambda
\end{equation}
below which spontaneous magnetization is possible.
We are (using the same approximation and following \cite[sec. 3.1]
{plischke}) able to construct an order parameter of the form
\begin{equation} \label{order}
2 \langle r \rangle - 1 = \pm (1 -{\beta_{c} \over \beta})^{1/2}.
\end{equation}
So, the exponent for the power law behavior of the order parameter is
 the same as for the hyperbolic tangent Brillouin function
(\ref{brill}) \cite[eq. (3.10)]{plischke}, that is the simple
fraction ${1 \over 2}$ --- as is not actually
the case for real ferromagnets.
(Lavenda and Florio \cite{florio} assert that the probability density
centered about the metastable state of the order parameter
for the mean-field theory of the kinetic Weiss-Ising model is
 an asymptotic distribution for the smallest value, rather than
a normal distribution, as generally assumed.)

The possibility of obtaining results analogous to those derived here,
for $n$-level systems ($n > 2$) should be investigated.
Such results could, then, be compared with those given by the particular
$n$-level form of the Brillouin function. In this regard,
it might be of some relevance to note that Brosseau \cite{bross96}
has recently shown that the polarization entropy of a stochastic
 radiation field
depends on ($n-1$) measures of the degree of polarization of the
field.

If one considers the pure states in a $2m$-dimensional complex Hilbert space,
endowed with the unitarily
 invariant integration measure \cite{jones}, then tracing over $m$ degrees
of freedom, one  induces a probability distribution on the two-level
standard complex quantum
systems taking the form of the Gibbs distribution (\ref{4}) with
$\beta = m -1$
\cite{page} (cf. \cite{derka}).

Boltzmann's principle --- $S = k \log \Omega$ + const ---relates the
entropy, $S$, to the logarithm of the thermodynamic probability,
$\Omega$, of a given state \cite[sec. II.4.1]{lav3}.
The question of whether this important
 principle has a meaningful application to
the structure functions considered here ($\Omega_{complex}, \Omega_{quat},
\Omega_{real} \equiv 1$, $\Omega_{class}$ 
 and $\Omega_{KMB}$) certainly merits investigation.
(We note that all but the last of these functions assume the value 1
for $E= \infty$, the logarithm of which is 0 equaling, as would
seem appropriate, the
{\it von Neumann}
 entropy, $ -\mbox{Tr} \rho \log \rho$, of a pure state --- that is,
one for which,
$\det {\rho} = 0$ or, equivalently, $r = 1$ or $E = \infty$ (cf. (\ref{2})).
However, the von Neumann entropy of the fully mixed state, corresponding
to $r = 0$ or $E=0$, is $\log 2$, not $-\infty$, as a naive application
of the principle would, then, give.)

White \cite{white} has developed a {\it density matrix} formulation for
quantum renormalization groups and applied it to Heisenberg chains.
(He shows that keeping the most probable eigenstates of the block density
matrix gives the most accurate representation of the system as a whole,
that is, the block plus the rest of the lattice.)
We have indicated here a different use of the density matrix concept in
studying spin systems, in line with the program expounded by Band and
Park. In their series of papers \cite{bp1,bp2,bp3,bp4}, they
 take exception to the Jaynesian strategy
of maximization of the von Neumann entropy
subject to constraints on expected values of observables \cite{jaynes,balian}
and suggest that rather one should estimate a Gibbs
distribution over the continuum or ``logical spectrum'' of possible
density matrices.
We should also point out, however, that Band and Park --- in contrast
to Lavenda \cite{lav1,lav2} and Tikochinsky and Levine \cite{tik} --- do
 not discuss the notion of an (improper or nonnormalizable) structure function
or density of states over the energy levels
(such as we encounter here with (\ref{3}), (\ref{quatstruct}),
(\ref{classstruct}),
 (\ref{KMBstruct}) and $\Omega_{real}(E) \equiv 1$), but solely
 that of a reparameterization-invariant
prior {\it probability} distribution over the convex set of
quantum systems (cf. \cite{slat7}).
 Within the framework of their
analysis, concerned with systems of arbitrary dimensionality (not simply
two-dimensional systems, as here), Band and Park
 did conclude \cite{bp3,bp4} that for the case of ``strong'' or
dynamical equilibrium, an axiom of ``data indifference'' was preferable
to one of ``state indifference''.

 In conclusion, let us state that we
 have  found counterparts (that is, the expected values $\langle r \rangle$
 of the radial
coordinate $r$ in the Bloch sphere-type representations of the two-level
systems) --- (\ref{expO}),
 (\ref{quatpolar}),
 (\ref{realexp}) and (\ref{polKMB}) --- to the hyperbolic
tangent Brillouin function (\ref{brill}), which are evidently
not subject to the methodological objections of Lavenda \cite{lav3}.
Of particular novelty has been the identification of the parameter
$\beta$ of the Gibbs canonical distributions introduced above with the effective
polarization temperature, and not, as is
conventional, the inverse thermodynamic temperature.

Further theoretical and empirical analyses pertaining to the potential
 applicability
to physical phenomena of the results reported here
 would, indeed, appear appropriate.

\acknowledgments

I would like to express appreciation to the Institute for
Theoretical Physics at UCSB for computational and other
technical support in this research
and to Christian Krattenthaler for comments on an early draft.


\begin{references}
\bibitem{bp1} J. L. Park and W. Band, Found. Phys. 6, 157 (1976).
\bibitem{bp2} W. Band and J. L. Park, Found. Phys. 6, 249 (1976).
\bibitem{bp3} J. L. Park and W. Band, Found. Phys. 7, 233 (1977).
\bibitem{bp4} W. Band and J. L. Park, Found. Phys. 7, 705 (1976).
\bibitem{jaynes} E. T. Jaynes, Phys. Rev. 108, 171 (1957).
\bibitem{balian} R. Balian and N. L. Balazs, Ann. Phys. (NY) 179,
97 (1987).
\bibitem{park} J. L. Park, Found. Phys. 18, 225 (1988).
\bibitem{gyft} E. P. Gyftopoulos and E. \c{C}ubuk\c{c}u, Phys. Rev. E 55, 3851 (1997).
\bibitem{slater} P. B. Slater, Phys. Lett. A 171, 285 (1992).
\bibitem{bross} C. Brosseau and D. Bicout, Phys. Rev. E 50, 4997 (1994).
\bibitem{tusz} J. A. Tuszy\'{n}ski and W. Wierzbicki,
Amer. J. Phys. 59, 555 (1991).
\bibitem{schlogl} F. Schl\"{o}gl, {\it Probability and Heat}
(Friedr. Vieweg, Braunschweig, 1989).
\bibitem{gold} E. V. Goldstein, Yu. Sadaui, and V. M. Tsukernik,
Phys. Rev. E 47, 3749 (1993).
\bibitem{lav3} B. H. Lavenda, {\it Thermodynamics of Extremes}
(Albion, West Sussex, 1995).
\bibitem{brezin} E. Br\'ezin, D. J. Gross, and C. Itzykson,
Nucl. Phys. B 235, 24 (1984).
\bibitem{abrahams} E. Abrahams and F. Keffer, in {\it McGraw-Hill
Encyclopaedia of Science and Technology} (McGraw-Hill, New York, 1977),
vol. 13, p. 96.
\bibitem{lee} K.-C. Lee, Phys. Rev. E 53, 6558 (1996).
\bibitem{boh} J. Bohman and C.-E. Fr\H{o}berg, Math. Comput. 58, 315 (1992). 
\bibitem{bal2} R. Balian, {\it From Microphysics to Macrophysics: Methods
and Applications of Statistical Physics} (Spinger, Berlin, 1991).
\bibitem{stanley} H. E. Stanley, {\it Introduction to Phase Transitions
and Critical Phenomena} (Oxford, New York, 1971).
\bibitem{als} B. N. Al-Saqabi, S. L. Kaller, and H. M. Srivastava,
J. Math. Anal. Appl. 159, 361 (1991).
\bibitem{christ} C. Christodoulides, Phys. Stat. Sol. (a)
109, 611 (1988).
\bibitem{mand} B. Mandelbrot, Phys. Today 42:1, 71 (1989).
\bibitem{lav1} B. H. Lavenda, Intl. J. Theor. Phys. 27, 451 (1988).
\bibitem{lav2} B. H. Lavenda, {\it Statistical Physics}
(Wiley, New York, 1991).
\bibitem{tik} Y. Tikochinsky and R. D. Levine, J. Math. Phys.
25, 2160 (1984).
\bibitem{kratt} C. Krattenthaler and P. B. Slater, {\it Asymptotic
Redundancies for Universal Quantum Coding}, Los Alamos preprint
archive, quant-ph/9612043 (1996).
\bibitem{braun2} S. L. Braunstein and G. J. Milburn, Phys. Rev. A
51, 1820 (1995).
\bibitem{ferguson} T. S. Ferguson, Ann. Statist. 1, 209 (1973).
\bibitem{bach} A. Bach, {\it Indistinguishable Classical Particles}
(Springer, Berlin, 1997).
\bibitem{braun} S. L. Braunstein and C. M. Caves, Phys. Rev. Lett.
72, 3439 (1994).
\bibitem{petzsud} D. Petz and C. Sudar, J. Math. Phys. 37, 2662 (1996).
\bibitem{petz3} D. Petz, Lin. Alg. Applic. 244, 81 (1996).
\bibitem{PETZ} D. Petz, J. Math. Phys. 35, 780 (1994).
\bibitem{slatjpa} P. B. Slater, J. Phys. A  29, L601 (1996).
\bibitem{clarke} B. S. Clarke and A. R. Barron, J. Statist. Plann. Inference
41, 37 (1994).
\bibitem{bied} L. C. Biedenharn and J. D. Louck, {\it Angular Momentum
in Quantum Physics} (Addison-Wesley, Reading, 1981).
\bibitem{pauncz} R. Pauncz, {\it Spin Eigenfunctions: Construction and Use}
(Plenum, New York, 1979).
\bibitem{farhi} E. Farhi and J. Goldstone, Ann. Phys. (NY) 192, 368 (1989).
\bibitem{mandel} M. J. Manfra {\it et al}, Phys. Rev. B 54, R17327 (1996).
\bibitem{chak} T. Chakraborty and P. Pietil\"{a}inen, Phys. Rev. Lett.
76, 4018 (1996).
\bibitem{man} M. J. Manfra {\it et al}, Phys. Rev. B 54, R17327 (1996).
\bibitem{riddle} T. W. Riddle {\it et al}, J. Vac. Sci. Technol.
15, 1686 (1978).
\bibitem{abram} {\it Handbook of Mathematical Functions}, edited by
M. Abramowitz and I. A. Stegun (Dover, New York, 1972).
\bibitem{frenzen} C. L. Frenzen, SIAM J. Math. Anal. 23, 505 (1992).
\bibitem{frenzen2} C. L. Frenzen, SIAM J. Math. Anal. 18, 890 (1987).
\bibitem{paris} R. B. Paris and A. D. Wood, J. Comput. Appl. Math.
41. 135 (1992).
\bibitem{adler} S. L. Adler, {\it Quaternionic Quantum Mechanics and
Quantum Fields} (Oxford, New York, 1995).
\bibitem{slaterjmp} P. B. Slater, J. Math. Phys. 37, 2682 (1996).
\bibitem{fivel} D. I. Fivel, Phys. Rev. A 50, 2108 (1994).
\bibitem{davies} A. J. Davies and B. H. J. McKellar,
Phys. Rev. A 46, 3671 (1992).
\bibitem{brumby} S. P. Brumby and G. C. Joshi, Chaos, Solitons 
and Fractals 7, 747 (1996).
\bibitem{beghian} L. E. Beghian, Nuovo Cim. B 107, 841 (1992).
\bibitem{D'Ariano} G. M. D'Ariano and M. G. A. Paris,
Phys. Rev. A 48, 4039 (1993).
\bibitem{lav6} B. H. Lavenda, Intl. J. Theor. Phys. 34, 605 (1995).
\bibitem{bern} J. M. Bernardo and A. F. M. Smith, {\it Bayesian Theory}
(Wiley, New York, 1994).
\bibitem{slat7} P. B. Slater, {\it Noninformative Priors for Quantum
Inference}, Los Alamos preprint archive, quant-ph/9703012 (1997).
\bibitem{petztoth} D. Petz and G. Toth, Lett. Math. Phys.
27, 205 (1993).
\bibitem{petzhas} D. Petz and H. Hasegawa, Lett. Math. Phys.
38, 221 (1996).
\bibitem{balescu} R. Balescu, {\it Equilibrium and Nonequilibrium
Statistical Mechanics} (Wiley, New York, 1975).
\bibitem{plischke} M. Plischke and B. Bergersen, {\it Equilibrium
Statistical Physics} (World Scientific, Singapore, 1994).
\bibitem{florio} B. H. Lavenda and A. Florio, Intl. J. Theor. Phys.
31, 1455 (1992).
\bibitem{bross96} C. Brosseau, Optik 104, 21 (1996).
\bibitem{jones} K. R. W. Jones, Ann. Phys. (NY) 207, 140 (1991).
\bibitem{page} D. N. Page, Phys. Rev. Lett. 71, 1291 (1993).	
\bibitem{derka} R. Derka, V. Bu\v{z}ek, and G. Adam, Acta Phys. Slov.
46, 355 (1966).
\bibitem{white} S. R. White, Phys. Rev. B 48, 10345 (1993); 
Phys. Rev. Lett. 69, 2863 (1992).
\end{references}
\end{document}